\documentclass{ws-ijmpe}

\usepackage{graphicx}% Include figure files
\usepackage{dcolumn}% Align table columns on decimal point
\usepackage{bm}% bold math

\begin{document}

\markboth{A. ULACIA REY}{GAUSS-BONNET CONTRIBUTIONS TO THE ENTROPY OF EXTREMAL BLACK HOLES IN THE GAUGE-GRAVITY SECTOR}

\catchline{}{}{}{}{}
%%%%%%%%%%%%%%%%%%%%%%%%%%%%%%%%%%%%%%%%%%%%%%%%%%%%%%%%%%%%%%%%%%%%

\title{GAUSS-BONNET CONTRIBUTIONS TO THE ENTROPY OF EXTREMAL BLACK HOLES IN THE GAUGE-GRAVITY SECTOR}

\author{\footnotesize A. ULACIA REY.}

\address{Department of Theoretical Physics, ICIMAF, Calle 15 No-309 esq-E, Vedado.\\
La Habana, Plaza cp-10400, Cuba.\\ alain@icmf.inf.cu}

% \author{SECOND AUTHOR}
% 
% \address{Group, Laboratory, Address\\
% City, State ZIP/Zone, Country\\
% second\_author@group.com}

\maketitle

\begin{history}
\received{(received date)}
\revised{(revised date)}
%\accepted{(Day Month Year)}
%\comby{(xxxxxxxxxx)}
\end{history}

\begin{abstract}
Using the Sen's mechanism we calculate the entropy for an $AdS_{2}\times S^{d-2}$ extremal and static
black hole in four dimensions, with higher derivative terms that comes from a three parameter non-minimal
Einstein-Maxwell theory. The explicit results for Gauss-Bonnet in the gauge-gravity sector are shown. 
\end{abstract}

\section{Introduction}

There are several forms of introduce higher derivative terms taking into account the gauge-gravity sector.
The standard way of perform the gauge-gravity interaction can be done in a minimal or non-minimal coupled picture.
However, the gauge sector can be added in a linear or non-linear form. Finally, the gravitational sector
can be added through a second order gravity or by terms of higher order of gravity. The more easy gauge-gravity
theory is the Einstein-Maxwell\cite{MTW} which is non-minimally coupled, is linear in the gauge sector and is of
second order in derivative in the gravity sector. In this case we can find solutions like are the Reissner-Nordstrom (RN) black
hole (BH) solution, and different models with electromagnetic fields\cite{RMW}.

 There are also other interesting solutions for minimally coupled gauge-gravity theory with non-linear gauge sector,
and second order in gravity. They are the well known Born-Infeld models\cite{Born_Infeld}, the Heisenberg-Euler
\cite{Hei_Euler} and more recently the Ayon-Beato and Garcia \cite{Eloy} solution.

The problem of the phenomenological introduction of non-minimal terms in the electrodynamic equations has been studied by
Hehl and Obukhov \cite{Hehl_Obukhov}. Some modifications of Maxwell equations has been done by Drummond and Hathrell
\cite{Drummond_Hathrell} from the one-loop corrections of quantum electrodynamics (QED) in curved space-time. 

In the following we will restrict our analysis to a coupling between gravity and electromagnetism in four
dimensions with the restriction of work with second order field equations. This kind of coupling is called the 
Gauss-Bonnet (GB) coupling (or GB solution) in a gauge-gravity sector and it allows only one (independent) non-minimal
coupling term in the action\cite{Horndeski_2}. These modifications of the Einstein-Maxwell theory arises from higher
dimensional Kaluza-Klein and string theory. In special, from a gravitational theory in five dimensions, via Kaluza-Klein
dimensional reduction, separating the quartic terms in the Maxwell field\cite{Buchdahl_Muller}. The dimensional reduction
of the Dirac action minimally coupled to gravity in five dimension leads a non-minimal coupling between Dirac field and
the electromagnetic field \cite{Pauli_Souriau}.

In other context, the entropy formalism\cite{Ferrara_1,Ferrara_2} has gained great importance because it has opened a
new road for find the entropy of extremal BHs with $AdS_{2}\times S^2$ topology. In the last years the Sen mechanism\cite{Sen:2007qy}
has been applied in several cases with nice results\cite{Examples_EF_1,Examples_EF_2,Examples_EF_3}.
For typical BPS BHs theses results for the entropy coincide with the statistical equation, the logarithm of the number
of BPS micro-states. Such coincidences is one of the most important contributions of the string theory in the last years.

In the article \cite{Ulacia} we calculated the entropy of extremal BHs using this mechanism and taking into account the complete
set of Riemann invariants, with a minimal coupling between gravity and gauge. Thus when we work with invariants of higher order
that comes from tensors as $R_{ab},R_{abcd},C_{abcd},S_{ab}$\footnote{They are respectively, the Ricci tensor, the Riemann tensor,
the Weyl tensor, the Trace-free tensor and more ahead the Maxwell tensor $F_{ik}$.} then we can say that working on the sector of pure
gravity, although the gauge group $U(1)$ could be present. Therefore if we introduced gauge invariants of higher order like can be
$(F^{ab}F_{ab})^2,F^{a}\,_{b}F^{b}\,_{c}F^{c}\,_{k}F^{k}\,_{a},...$ we are referring to the pure gauge sector, then the gauge-gravity
sector must be any combination of invariants, like could be $RF^{2},g_{ab}F^{ab},..$. For simplicity, other invariants of covariant
derivation like ${R_{a}}^{\,b}\,_{;c}\,{R_{b}}^{\,a;}\,^{c},{S_{a}}^{\,b}\,_{;c}\,{S_{b}}^{\,a;}\,^{c},...$ or another will not be
considered. This article constitutes our first step in introducing invariants from the gauge-gravity sector and obtain exact solutions
of this. All calculations were carried out with the GR-Tensor package running on the algebraic Maple program. 

This paper is organized in 4 Sections. First in Section \ref{Non-minimal_theory} we consider the non-minimal coupled theory.
Then in Section \ref{Sen_Mech} we apply the Sen mechanism on the GB theory. Finally in Section \ref{Conclu} a brief conclusion is
displayed. 
            
\section{Non-minimal coupling theory with linear coupling in curvature}
\label{Non-minimal_theory}

 A special case of non-minimal coupling is the theory with a restricts Lagrangian. It has a coupling between the electromagnetism
and the metric which is linear in the curvature. Such action takes the form \cite{Balakin_1,Balakin_2},

\begin{equation}
  S= \int{ dx^{4}\,\sqrt{-g}\,\left(\frac{R}{16\pi G_{4}}-\frac{F^{2}}{4}+\frac{w}{2}\chi^{ikmn}F_{ik}F_{mn}\right)}.
\label{Action_non_minimal}
\end{equation}
where $\chi^{ikmn}$, is the susceptibility tensor\footnote{All the other quantities are commonly well-known,
$R$ is the Ricci scalar,$G_{4}$ the Newton constant and $w$ the gauge-gravity coupling constant.}.
The physics reason for the last term is that the induction tensor $H^{ik}$, and the Maxwell tensor $F_{mn}$ are linked by the
linear law,
\begin{equation}
 H^{ik}=F^{ik}+\chi^{ikmn}F_{mn},
\end{equation}
as the magnetization tensor is defined by, 
\begin{equation}
 4\pi M^{ik}=H^{ik}-F^{ik}\equiv \chi^{ikmn}\,F_{mn},
\end{equation}
In the standard electrodynamics\cite{Landau} the coefficients $\chi^{ikmn}$ form the non-minimal susceptibility tensor.
The susceptibility tensor has the form,
\begin{equation}
 \chi^{ikmn}=\frac{q_{1}\,R}{2}(g^{im}g^{kn}-g^{in}g^{km})+\frac{q_{2}}{2}(R^{im}g^{kn}-R^{in}g^{km}+
R^{kn}g^{im}-R^{km}g^{in})+q_{3}\,R^{ikmn}. \label{suscep_tensor}
\end{equation}
the parameter $q_{1},q_{2}$ and $q_{3}$, are in general arbitrary. They can be chosen phenomenologically or by other way. 
At the same time, they introduce cross-terms which describe non-minimal interactions of the electrodynamics and gravitational
fields. For example, the Drummond-Hathrell\cite{Drummond_Hathrell} constraint modified the Maxwell equations from one-loop 
correction of QED in curved space-time, in this case the coupling constants can be written as, $w=1,\,2q_{1}-q_{3}=0,\,
13q_{1}+q_{2}=0,\,q_{1}=-{\alpha \lambda^{2}_{e}}/{180\pi},$ where $\alpha$ is the fine structure constant and
$\lambda_{e}$ is the Compton wavelength of the electron. If $q_{1}=q_{2}=q_{3}=w=0$, in Eq.~(\ref{suscep_tensor}) the RN
solution is displayed.

The Gauss-Bonnet model\cite{Horndeski,Muller} (in the context of gauge-gravity sector) is obtained when is request the
proportionality between the susceptibility tensor and the double-dual Riemann tensor $\chi_{ikmn}=q\,R_{ikmn}^{\ast\,\ast}$. 
Here the Einstein-Maxwell coupling equations are second order in the derivatives, for that reason is called Gauss-Bonnet
model, it is also an one-parameter model with $q_{1}=q_{3}=-1,q_{2}=2$ and $w=q$. 

In general these cross-terms in the Lagrangian represent the interactions between gravity and electromagnetic forces at
non-minimal level, therefore they means that the velocity of the gauge-gravity waves differ from the speed of the light
in the vacuum. 

\section{Applying the Sen mechanism}  
\label{Sen_Mech}
The Sen mechanism has allowed to compute the entropy for extremal BH in theories with higher derivative terms.
There are several examples that prove the efficiency of this mechanism.
%\cite{Examples_EF_1,Examples_EF_2,Examples_EF_3}.

The more general metric for an extremal and stationary BH with $AdS_{2}\times S^{2}$ topology in 4-dimensions is,

\begin{eqnarray}
 ds^{2}&=& v_{1}\left(-r^2\,dt^2+\frac{dr^2}{r^2}\right)+v_{2}(d\theta^2+sin^{2}\theta d\phi^2),\\
 F_{rt}&=& e, \qquad F_{\theta \phi}=p\,sin\theta/4\pi.  
\end{eqnarray}
where $v_{1},v_{2}$ are related with the geometry of the BH throat, $e$ and $p$ are related with the BH charges.
Is important to point out that we are assuming our gauge-gravity theory in connexion with the $AdS_{2}\times S^{2}$.
This means that our theory must contain the RN solution\footnote{If $w=0$ in Eq. \ref{Action_non_minimal} the RN solution
is obtained.} that at the extreme limit of larger charges, near the horizon of geometry could be connected with the
$AdS_{2}\times S^{2}$ topology and the associated isometry $SO(2,1)\times SO(3)$. Furthermore, is natural postulate that the
effects of added higher derivative terms from the gauge-gravity sector will not destroy the symmetries near the horizon of geometry
\cite{Sen:2007qy,James2}. Thus let us define $\mathcal{E}$ (the entropy function) like,
\begin{equation}
 \mathcal{E}(\vec{v},\vec{e},\vec{q},\vec{p})=2\pi (e_{i}q_{i}-f(\vec{v},\vec{p},\vec{e})).
\end{equation}    
where $f$ is the Lagrangian density evaluated for the near horizon of geometry. $\mathcal{E}$ contains all the information near the
horizon of the geometry. Thus all the extremal conditions of the parameter determine the equations of motion and the conservation
near the horizon of geometry. The value of $\mathcal{E}$ on the horizon of geometry, is the entropy of the BH. 

If we calculate these quantities for a gauge-gravity non-minimally coupled theory in the Gauss-Bonnet case we get for the entropy
function,
\begin{equation}
 \mathcal{E}=-\left(\frac{16\pi^{2}e^2}{v_{1}}+\frac{p^2}{v_{2}}\right)w+2\pi eq-\frac{4\pi^{2}e^{2}v_{2}}{v_{1}}+
\frac{v_{1}p^{2}}{4v_{2}}-\frac{\pi (v_{1}-v_{2})}{G_{4}}. \label{Entr_Func}
\end{equation}
 The equations of motion are determined by extremizing the entropy function,
\begin{equation}
\frac{\partial\mathcal{E}}{\partial\,{v_{j}}}=\frac{\partial\mathcal{E}}{\partial\,{e}}=0,\qquad \hbox{with} \qquad j=1,2. 
\end{equation}
Therefore the equations of motion near the horizon of geometry can be written as follows,
\begin{eqnarray}
\frac{-16\pi^{2}e^{2}w}{v^{2}_{1}}-\frac{p^{2}}{4v_{2}}+\frac{\pi}{G_{4}}-\frac{4v_{2}e^{2}\pi^{2}}{v^{2}_{1}}&=&0,\\
\frac{p^{2}w}{v^{2}_{1}}-\frac{v_{1}p^{2}}{4v^{2}_{2}}+\frac{\pi}{G_{4}}-\frac{4\pi^{2}e^{2}}{v_{1}}&=&0,\\   
\frac{-ew}{v_{1}}+\frac{q}{16\pi}-\frac{v_{2}e}{4v_{1}}&=&0.
\end{eqnarray}
Thus when we solve this system of equations we get two set of solutions. One of them can be discarded because it does not contain
the RN solution, hence we get of extremal conditions for,
 \begin{eqnarray}
   v_{1}&=& v_{2}+4w, \qquad \qquad e=\frac{q}{4\pi},  \label{Sol_GB_v1_e_q} \\
   v_{2}&=& \frac{(G_{4}(p^2+q^2)-16\pi w)}{8\pi}\left[1+\sqrt{1+\frac{4^{3}\pi w G_{4}p^2}{16\pi w-
G_{4}(p^2+q^2)} }\,\right].\nonumber  
 \end{eqnarray}
Substituting this into the expression for $\mathcal{E}$ in (\ref{Entr_Func}) we get,
\begin{equation}
 S_{GB}=\frac{\pi v_{2}}{G_{4}}-\frac{p^2\,w}{v_{2}}. \label{Sol_Entr_GB}
\end{equation}
Clearly, the Eqs.(\ref{Sol_GB_v1_e_q})-(\ref{Sol_Entr_GB}) reproduces the RN solution when $w=0$.
But from Eq. (\ref{Sol_GB_v1_e_q}), we can see that the introduction of the higher derivative
terms in the action makes a tiny change in the geometry of the BH throat. Normally, under this
limit, near the horizon of geometry, the coupling constant must be inverse to the powers of the
electric and magnetic charges. Therefore, the correction to the entropy of RN is small in Eq.
(\ref{Sol_Entr_GB}). This GB solution for entropy is contained in the gauge-gravity sector and
introduce the same effect than the typical GB in the sector of pure gravity. Both modify the
entropy of Bekenstein-Hawking. Note that the GB solution in the sector of pure gravity takes
form\cite{Morales:2006gm,Ulacia},
\begin{eqnarray}
 v_{1}&=& v_{2}, \qquad e=\frac{q}{4\pi}, \qquad v_{2}= \frac{G_{4}}{4\pi}(p^2+q^2), \\
 S^{grav}_{GB}&=&\frac{\pi v_{2}}{G_{4}}+64\pi^{2}\alpha. \label{Entr_GB_PG}
\end{eqnarray}
where $\alpha$ is the coupling constant. This solution does not change the throat topology $(v_{1}=v_{2})$,
just the introduction of a cosmological constant can do that\cite{Ulacia}. As we supposed, near the horizon of geometry
the coupling constant is inverse to the BH charge, then its contribution in Eq .(\ref{Entr_GB_PG}) is also small. 

\section{Conclusions}
\label{Conclu}
 The Sen mechanism let us to find exact results for the entropy of extremal BH in four dimensions with $AdS_{2}\times S^{2}$
topology, taking into account terms of higher derivatives in the sector of gauge-gravity of a GB theory.
These results were compared with the similar ones in the sector of pure gravity of the GB theory. In both cases and with
small contributions, the area law is modified.    
\section*{Acknowledgements}
The author acknowledge thanks the Office of External Activities of ICTP for its support through NET-35.
Also thank to CLAF (Latin-American Center for Physics) and Cuban Physics Society for their contributions
and supports in the organization of the events STARS-2011 and SMFNS-2011.
\appendix

\end{document}